%% file: main.tex
\documentclass[9pt,journal]{IEEEtran}
\usepackage{authblk}
\usepackage[utf8]{inputenc}
\usepackage{caption}
\usepackage{graphicx}
\usepackage{subcaption}
\usepackage[dvipsnames]{xcolor}
\usepackage{gensymb}

\title{Reconfigurable Intelligent Surface-Assisted Backscatter Communication: A New Frontier for \\ Enabling 6G IoT Networks}

\author{Sarah~Basharat,~Syed~Ali~Hassan,~\IEEEmembership{Senior~Member,~IEEE,}~Aamir~Mahmood,~\IEEEmembership{Senior~Member,~IEEE,}\break Zhiguo Ding,~\IEEEmembership{Fellow,~IEEE,}~and~Mikael~Gidlund,~\IEEEmembership{Senior~Member,~IEEE}

\thanks{S. Basharat and S. A. Hassan  are with the School of Electrical Engineering
and Computer Science (SEECS), National University of Sciences and Technology (NUST), Islamabad 44000, Pakistan.}
\thanks{A. Mahmood and M. Gidlund are with Mid Sweden University, Sweden.}
\thanks{Z. Ding is with the School of Electrical and Electronic Engineering, The University of Manchester, Manchester M13 9PL, U.K.}%
}

\usepackage{amsmath}
\usepackage{amssymb}
\usepackage{cite}
\usepackage{multirow}
\usepackage{tabularx}
\newcolumntype{P}[1]{>{\centering\arraybackslash}p{#1}}
\newcolumntype{M}[1]{>{\centering\arraybackslash}m{#1}}
\begin{document}
\maketitle
\begin{abstract}
Backscatter Communication (BackCom), which is based on passive reflection and modulation of an incident radio-frequency (RF) wave, has emerged as a cutting-edge technological paradigm for self-sustainable Internet-of-things (IoT). Nevertheless, the contemporary BackCom systems are limited to short-range and low data rate applications only, thus rendering them insufficient on their own to support pervasive connectivity among the massive number of IoT devices. Meanwhile, wireless networks are rapidly evolving towards the smart radio paradigm. In this regard, reconfigurable intelligent surfaces (RISs) have come to the forefront to transform the wireless propagation environment into a fully controllable and customizable space in a cost-effective and energy-efficient manner. Targeting the sixth-generation (6G) horizon, we anticipate the integration of RISs into BackCom systems as a new frontier for enabling 6G IoT networks. In this article, for the first time in the open literature, we provide a tutorial overview of RIS-assisted BackCom (RIS-BackCom) systems. Specifically, we introduce the four different variants of RIS-BackCom and identify the potential improvements that can be achieved by incorporating RISs into BackCom systems. In addition, owing to the unrivaled effectiveness of non-orthogonal multiple access (NOMA), we present a case study on an RIS-assisted NOMA-enhanced BackCom system. Finally, we outline the way forward for translating this disruptive concept into real-world applications.

\end{abstract}
\section{Introduction}
\IEEEPARstart{T}{HE} Internet-of-Things (IoT) is envisaged as one of the key technology trends towards the development of intelligent Internet solutions for future sixth-generation (6G) systems~\cite{6G1}. The IoT paradigm targets seamlessly connecting massive power-limited sensor-like devices, with data sensing and transmission capabilities, for realizing diverse applications, such as environmental sensing, industrial automation, pervasive monitoring, intelligent transportation, smart farming, and smart cities. According to the Cisco report, “500 billion devices are expected to be connected to the Internet by 2030”~\cite{Cisco}. Nonetheless, despite the rapid research and evolution over the recent years, IoT is still at the germination stage, awaiting a large-scale deployment and a wide-range commercialization. In many IoT applications, powering the IoT devices is a major challenge, especially when battery maintenance is not feasible due to cost, inconvenience, and the size of the networks. In this context, energy harvesting is a promising solution for self-sufficient and self-sustainable operations of IoT networks.

Backscatter communication (BackCom), one of the energy harvesting techniques, has emerged as an energy-efficient solution for pervasive connectivity of power-limited wireless devices in an IoT network~\cite{Shahzeb_NOMA_BackCom_Mag}. BackCom enables the passive backscatter devices (BDs) to not only transmit information, through reflection and modulation of incident RF signals via intentional load impedance mismatch at the antenna, but also to harvest energy from the incident signals to sustain its operations. Thus, with this energy saving feature, BackCom achieves a green communication paradigm for IoT networks.

The architecture of BackCom systems has evolved over time with three different configurations, namely monostatic, bistatic, and ambient~\cite{BackCom_systems}. 

\textbf{Monostatic BackCom Systems:} In a monostatic BackCom system, such as RF identification (RFID), the carrier emitter (CE) and backscatter receiver (BR), are integrated as a single equipment, i.e., reader. The reader transmits the RF signal to the BD, which then loads its own information onto the incident signal and backscatters the modulated signal to the reader. Since the CE and BR are co-located, the modulated signal suffers from round-trip path loss. Consequently, a monostatic BackCom system is limited only to short-range applications.

\textbf{Bistatic BackCom Systems:} Unlike monostatic BackCom system, the CE and BR are spatially separated in the bistatic configuration, thus improving the flexibility of the system by allowing the optimal placement of CE. In addition, the bistatic BackCom system improves the performance in a cost-effective manner due to a less complex design than the monostatic counterpart. However, for long-range applications, the bistatic configuration requires the dedicated RF source to be in close vicinity of BDs in an interference-limited region.

\textbf{Ambient BackCom Systems:} Similar to bistatic configuration, the CE is also separated from the BR in an ambient BackCom system. However, instead of a dedicated RF source, it exploits the ambient RF sources, such as TV/cellular towers, Bluetooth, or WiFi access points, which reduces the cost and power consumption of a dedicated CE. Moreover, the ambient configuration improves the spectrum resource utilization since it utilizes the existing RF signals. However, in an ambient configuration, the reader suffers from direct-link interference from the ambient signals, which adversely affects the detection performance and consequently limits the transmission range. 

Despite the extensive research, there exist a fundamental limitation in signal coverage and data rate of BackCom systems, hindering the large-scale deployment. Therefore, alternative methods needs to be developed, which can support the ubiquitous connectivity among a massive number of devices, while adapting to the dynamic nature of the wireless channels. This makes it necessary to control and manipulate the electromagnetic waves to steer the signals in the desired directions,  thus paving a way towards the smart and reconfigurable propagation environments. Hence, in this regard, reconfigurable intelligent surfaces (RISs), also known as intelligent reflecting surfaces (IRSs), have come to the forefront to improve the  propagation conditions by passive signal reflections~\cite{IRS_Rui_Zhang,Marco_IRS}.

Specifically, an RIS is a planar array of numerous low-cost passive reflecting elements, each capable of inducing a phase-shift in the impinging signal. Through intelligent phase-shifts, the RIS reflected signal can be combined with direct link signals, either constructively to boost the received signal strength or destructively to attenuate the co-channel interference, hence improving overall system performance. These properties make RIS an ideal candidate to assist the BackCom systems to improve the network coverage, lifetime, and capacity. Thus, as a change on the network infrastructure level, the application of RISs into BackCom systems opens an entirely new realm for IoT. 

Inspired by the benefits that RIS can bring to BackCom systems, this article is a first effort to give a forward-looking vision of the role that RISs can play in enabling smart BackCom systems. The key contributions of this article are summarized as follows.
\begin{itemize}
    \item To develop an understanding of RIS in BackCom, we provide a background of the fundamental principle, hardware architecture, and major advantages of RIS technology. 
    \item To realize the smart BackCom, we introduce four different variants of RIS-assisted BackCom (RIS-BackCom), namely, RIS-assisted monostatic, bistatic, ambient, and phase-shift BackCom systems.
    \item We unleash the five potential improvements that can be achieved by integrating RISs into BackCom systems.
    \item Owing to the unrivaled effectiveness of non-orthogonal multiple access (NOMA), a novel case study is presented which unveils RIS-assisted NOMA-enhanced BackCom system as a potential candidate for 6G IoT networks.
    \item We identify some critical challenges and research opportunities to provide the way forward for realizing RIS-BackCom systems.
\end{itemize}

\section{RIS Technology: An Overview}
In this section, to shed light on the basics of RIS, we provide an overview of the RIS fundamental principle, hardware architecture, and major advantages.
\vspace{-0.1cm}
\subsection{Fundamental Principle}
RIS implementation is based on the concept of synthetically produced two-dimensional form of electromagnetic metamaterials, known as the metasurface, which is composed of a massive number of reflecting elements, called meta-atoms~\cite{IRS_Rui_Zhang}. The electromagnetic behavior of the metasurface is controlled by the geometrical properties, i.e, size, orientation, arrangement, etc., of the meta-atoms. However, owing to the stochastic nature of propagation channels, the wireless communication applications demand real time response of meta-atoms, which can be realized through electronic devices, functional materials, or mechanical actuation. 

\subsection{Hardware Architecture}
\begin{figure}[t]
\centering   
\includegraphics[width=0.5\textwidth]{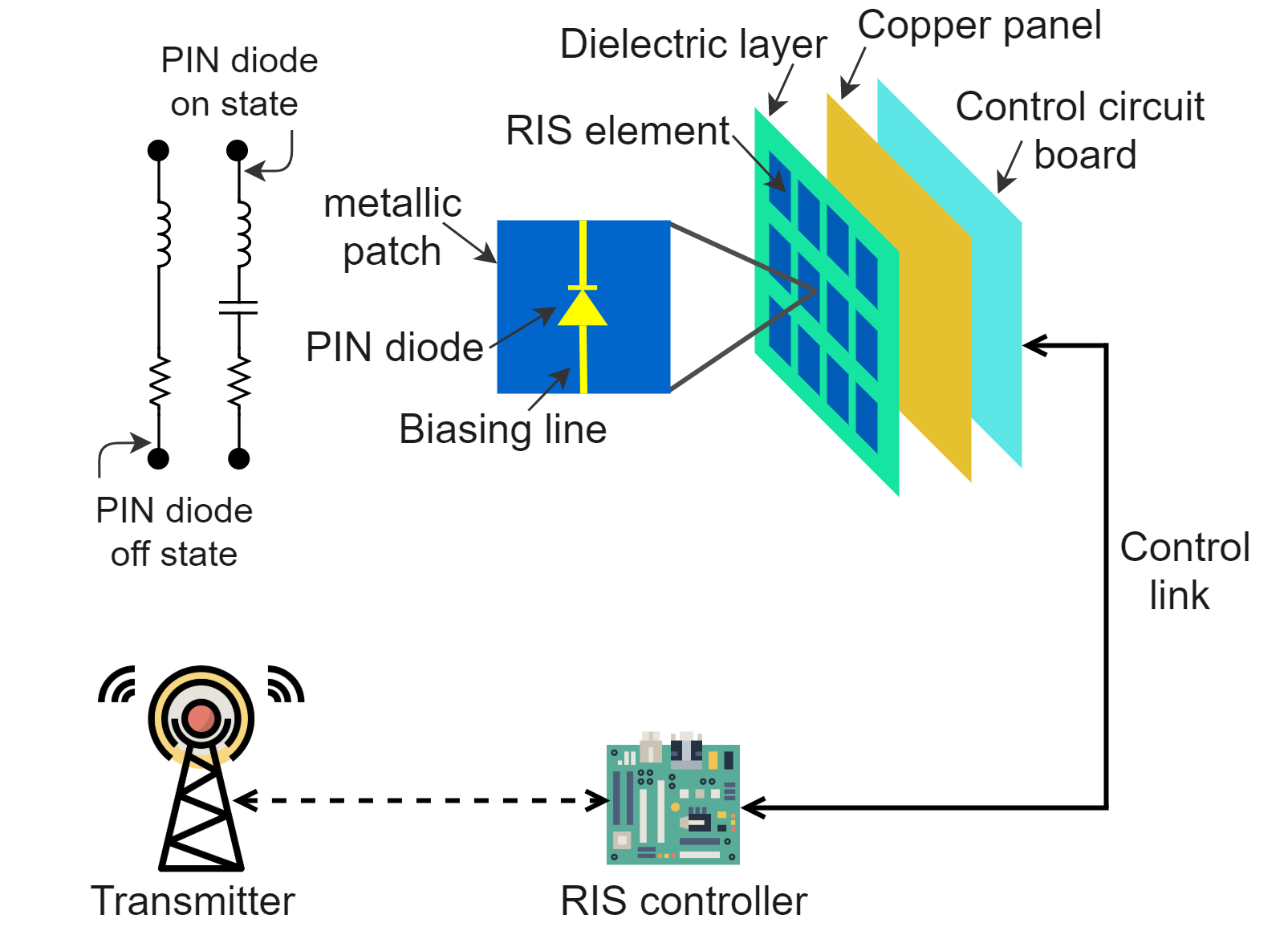}
\caption{Hardware architecture of RIS.}\label{fig.1}
\end{figure}

The hardware architecture of RIS comprises of two parts: the multi-layer planner surface and a smart controller~\cite{IRS_Rui_Zhang}. In a typical three-layered architecture, as shown in Fig.~\ref{fig.1}, the outer layer consists of a large number of elements printed on a dielectric substrate to directly interact with the incident signals. The middle layer is a copper panel, which avoids any signal leakage. The inner layer is a control circuit board, responsible for adjusting the reflection amplitudes and/or phase-shifts of each RIS element. The reconfiguration of RIS elements is managed by the smart controller, such as a field-programmable gate array (FPGA), which also acts as a gateway to communicate with the base station (BS) and/or other network components.

The structure of an individual element, embedded with a positive-intrinsic-negative (PIN) diode, is also shown in Fig.~\ref{fig.1}. By controlling the voltage via biasing line, the PIN diode can be switched between on and off states to realize the phase-shift of \(\pi\) in radians. Moreover, the RIS elements can also be equipped with low-cost dedicated sensors to enable sensing capabilities beneficial for wireless channel estimation.

\subsection{Major Advantages}
The RIS technology is not only conceptually appealing, but also offers several advantages for practical implementation. The RIS elements passively reflect the impinging signal, eliminating the requirement of RF chains and sophisticated signal processing. Hence, RIS can be implemented at a much low hardware cost and power consumption than active antenna counterparts. Moreover, different from conventional relays, RIS operates in a full-duplex mode (FD), free form self-interference and thermal noise. Architecturally, RIS is lightweight with conformal geometry and can be easily mounted on the ceilings, walls, and building facades. Additionally, the RIS-enhanced networks are compatible with the emerging communication technologies, such as non-orthogonal multiple access (NOMA), millimeter wave (mmWave) communication, unmanned aerial vehicles (UAVs), and BackCom~\cite{Marco_IRS}. These advantages make RIS a cutting-edge technology for realizing 6G wireless networks~\cite{Sarah_WCM_RIS}.

\section{Variants of RIS-BackCom}
\begin{figure*}[t]
\centering   
\includegraphics[width=0.98\textwidth]{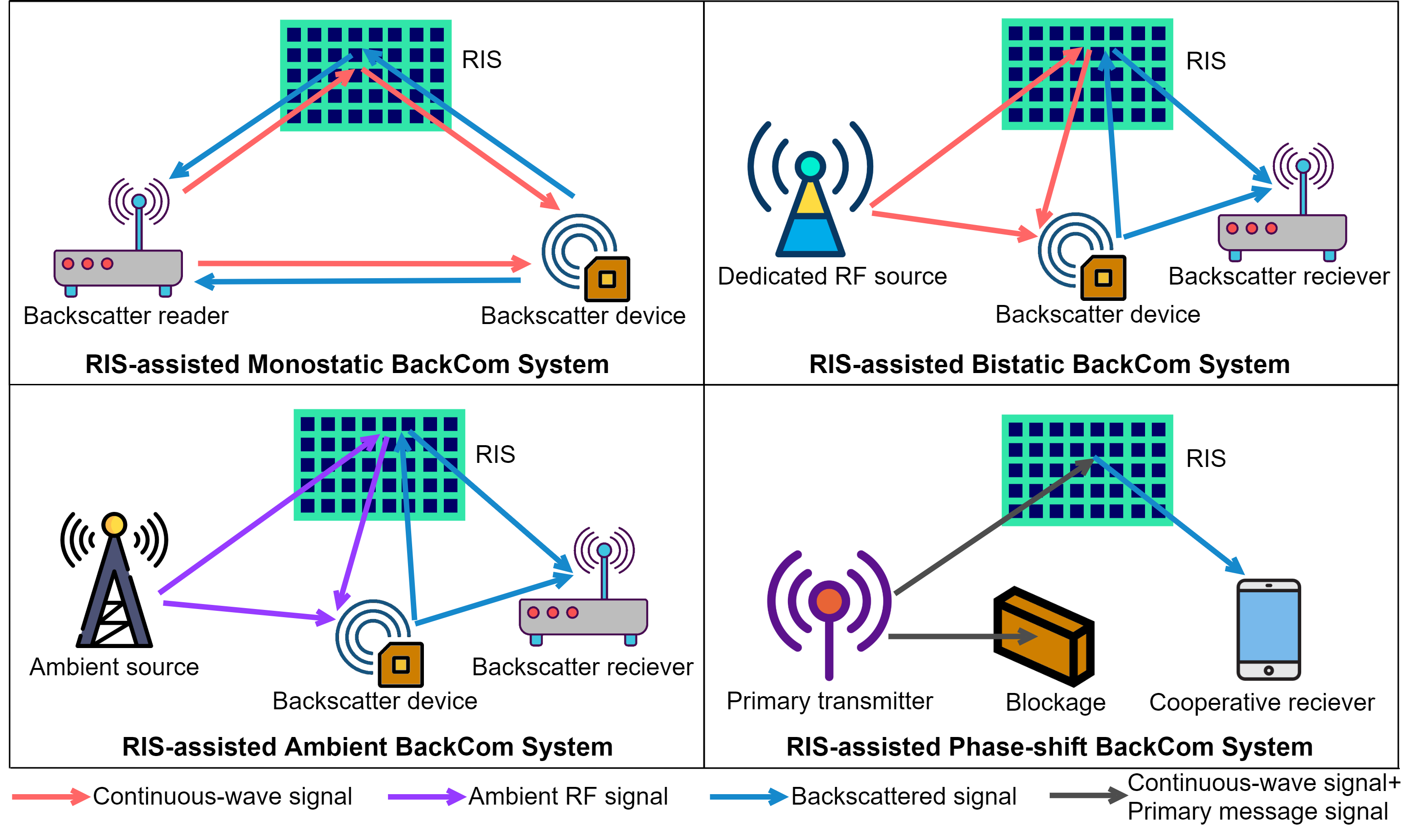}
\caption{Variants of RIS-BackCom}\label{fig.2}
\end{figure*}

RIS can be incorporated in BackCom systems to realize the future green and ubiquitous communication for IoT applications. To this end, in this section, we discuss the four different variants of RIS-BackCom, as shown in Fig.~\ref{fig.2}.

\subsection{RIS-assisted Monostatic BackCom System}
The communication in an RIS-assisted monostatic BackCom system can be split into two phases, the excitation phase and the backscattering phase. In the excitation phase, the reader sends the continuous-wave signal to the BD directly and via smart reflection from RIS. In the backscattering phase, the BD modulates the signal and then backscatters it towards the reader via both the direct and RIS reflected links. The intelligent phase-shifts induced at the RIS result in the improvement of the received signal-to-noise ratio (SNR), which can be mapped to extend the transmission ranges of BDs.

\begin{figure*}[t]
\centering   
\includegraphics[width=0.98\textwidth]{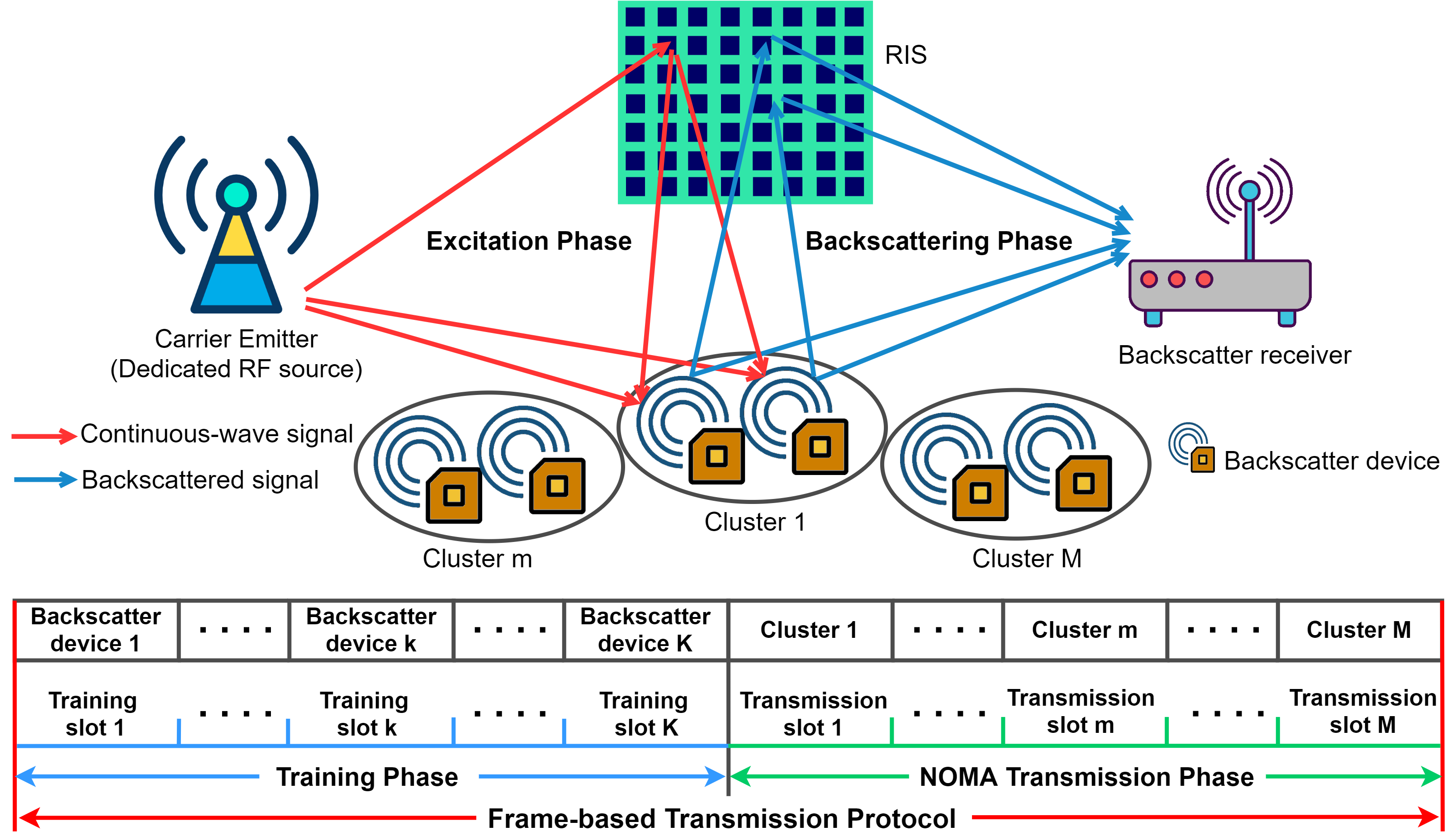}
\caption{An illustration of RIS-assisted NOMA-enhanced bistatic BackCom system.}\label{fig.3}
\end{figure*}

\subsection{RIS-assisted Bistatic BackCom System}
Similar to RIS-assisted monostatic BackCom system, in the bistatic configuration, the ability of RIS to reconfigure the wireless medium can be utilized to enhance the performance of both the forward (i.e., excitation) and backward (i,e., backscattering) links. The direct and RIS reflected signals are added coherently at both the BD and BR, resulting in favorable SNR scaling, which can be translated to reduce the transmit power requirement, extend the transmission ranges, and enhance the device lifetimes. Moreover, the application of RIS to a bistatic BackCom system allows the flexible placement and deployment of BDs, providing an additional degree of freedom.

\subsection{RIS-assisted Ambient BackCom System}
In an RIS-assisted ambient BackCom system, the BD conveys information on the top of an already modulated ambient signal, received directly from the ambient source and via reflection from RIS. In the  ambient configuration, the ability of RIS to focus the signals in the desired direction can be utilized to mitigate the direct-link interference from the ambient signals, which can improve the detection performance at low complexity. Moreover, similar to RIS-assisted monostatic and bistatic configurations, incorporating RIS in an ambient BackCom system can enhance the coverage range, data rate, and energy harvesting capability of BDs at reduced transmit power requirements.

\subsection{RIS-assisted Phase-shift BackCom System}
In an RIS-assisted phase-shift BackCom system, the RIS substitutes the conventional BDs to relay the primary message signal and convey the secondary information, modulated over the CW signal, to the cooperative receiver~\cite{IRS_phase_shift_BackCom}. In this configuration, RIS introduces an additional phase-shift for the secondary information while reflecting the primary message signal and unmodulated carrier. The cooperative receiver, with a sharp band-pass filter, receives both the secondary and primary signals and decodes them sequentially. Hence, with effective power splitting among the primary message signal and unmodulated carrier, the RIS-assisted phase-shift BackCom system can yield better spectrum efficiency than a conventional BackCom system.

\section{RIS-BackCom: Potential Improvements}

In this section, we elaborate on the potential improvements that can be achieved by integrating RIS in BackCom systems.

\subsection{Reconfigured Channel Conditions}
In conventional BackCom systems, both the continuous-wave, or ambient RF signal, and the backscattered signal undergo various propagation phenomena, such as reflection, diffraction, and scattering, which result in multiple copies of the signals. These copies usually arrive out-of-phase at the BD, in the excitation phase, and the BR, in the backscattering phase, which produce significant distortions in the received signals, thus limiting the performance of BackCom systems. However, the limitations of conventional BackCom systems can be compensated via RIS. Through smart reflections, RIS can mitigate the negative effects of electromagnetic radiations, thereby improving the system performance.

RIS can establish virtual line-of-sight (LoS) links to improve the overall channel gains. For instance, in a dense environment, such as an industrial IoT network, RIS deployed at the optimal location can bypass the environmental obstacles to establish communications. Such an application scenario is more evident in high-frequency bands, i.e., millimeter waves (mmWaves) and Terahertz (THz), which are vulnerable to signal blockage and attenuation. In this regard, the authors in~\cite{IRS-MBC_SEP} analyzed the reliability performance of an RIS-assisted monostatic BackCom system, where the direct link between the tag and the reader is assumed to be blocked. The RIS-BackCom can achieve high reliability than a relay-assisted BackCom at a moderate SNR with a massive number of reflective elements.

\subsection{Enhanced Transmission Range}
Despite the research achievements, the conventional BackCom systems offer inherently limited range, in the order of a few meters, which is a major hurdle preventing their large-scale deployment. Specifically, monostatic systems suffer from round trip path loss; bistatic systems require the CE to be placed very close to BDs; and ambient systems experience direct-link interference, thus restricting the BackCom to short-range low date rate applications only. The conventional solutions to overcome the short coverage impairments include relay-aided and multi-antenna backscattering. However, such techniques incur high infrastructure cost, complexity, and power consumption. Fortunately, RIS-BackCom system can boost the range and coverage of BDs in an energy-efficient and cost-effective manner. 

RIS can be installed at the edge of receiver's coverage zone to enhance the transmission range and cover the BDs located in the dead zones, i.e., regions
with no signal reception. As a result, BDs connectivity and coverage can be enhanced with the aid of RIS. For the coverage enhancement of ambient BackCom systems, the authors in~\cite{IRS-AmBC_Short_Range} proposed a novel RIS-assisted ambient BackCom technique over ambient orthogonal-frequency-division-multiplexing (OFDM) subcarriers. The proposed technique benefits from multipath gains reflected from the RIS elements 
in terms of enhanced received signal power and consequently coverage.

\subsection{Improved Energy Efficiency}
The most attractive aspect of RIS technology, from an energy consumption standpoint, is the ability to amplify and forward the impinging signal without employing any power-hungry source, but rather by appropriately designing the RIS phase-shifts to constructively combine the reflected signals. Hence, a BackCom system can benefit from the power gains realized through RIS passive reflections to achieve higher performance gains with less transmit power, which can significantly improve the energy efficiency of RIS-BackCom systems.

As an exploratory work towards an energy-efficient design for future BackCom systems, the authors in~\cite{MBC_BBC_Tx_min} proposed a novel setup for a bistatic BackCom system, where an RIS is deployed close to the CE to assist the communication between the BD and the reader. The joint optimization of RIS phase-shifts and transmit beamforming vector at the CE can significantly reduce the transmit power consumption, while guaranteeing a required BackCom performance. The simulation results reveal the transmit power reduction of $3.5$ dB by deploying an RIS with $49$ reflecting elements. Moreover, further reduction in transmit power can be achieved by increasing the reflecting elements. In addition, the authors extended the proposed design for transmit power minimization to a multi-device scenario in~\cite{IRS-BBC_multi_user}, resulting in a similar reduction in transmit power, which can be translated to improve the link budget and transmission ranges. However, currently, there is a paucity of research contributions to investigate the trade-offs between the transmit power consumption and throughput of RIS-BackCom systems for maximizing the energy efficiency, which is worth exploring in future.

\subsection{Better Detection Performance}
Amongst the three contemporary BackCom configurations, ambient BackCom is the most energy-efficient and cost-effective solution for the pervasive networking of massively deployed low-power IoT devices. Nonetheless, the robustness of ambient BackCom systems is a prominent challenge that needs to be addressed. Specifically, in ambient BackCom systems, the backscattered signal experiences direct-link interference from the ambient signals, thus rendering poor detection performance at the reader. Although the conventional methods, such as channel estimation and modification of BDs, can improve the detection performance, however, at the cost of high implementation complexity.

To tackle the aforementioned limitation, the ability of RIS to steer the signals in different directions can be exploited to suppress the direct-link interference, thus improving the detection performance at reduced complexity. In this regard, the authors in~\cite{IRS-AmBC_Deep_Learning} proposed deep reinforcement learning (DRL) based approach, namely the deep deterministic policy gradient (DDPG) algorithm, to optimize the performance of an RIS-assisted ambient BackCom system in the absence of channel state information (CSI). The simulated results demonstrate the better detection performance of the proposed design over the conventional non-RIS based designs with full CSI knowledge. 

\subsection{High Energy Harvesting}

Energy harvesting is an attractive technique towards the battery-free operation of IoT networks. However, currently, the typical energy conversion efficiency in BackCom is usually less than $20$\%. Such a poor energy harvesting performance can severely affect the self-sustainable operation of future IoT networks. 
To address this challenge, RIS can be applied to BackCom systems to improve the total received power by a coherent combination of reflected signals. Thus, enabling the BDs to harvest energy from both the direct and RIS reflected signals, consequently, improving the total harvested energy to sustain the long-term operation of IoT networks.

Despite the exploratory works~\cite{IRS-MBC_SEP,IRS-AmBC_Short_Range,MBC_BBC_Tx_min,IRS-BBC_multi_user,IRS-AmBC_Deep_Learning}, the performance of RIS-BackCom systems in the context of energy harvesting has not been analyzed yet. Therefore, it is necessary to design efficient frame-based transmission protocols for RIS-BackCom systems to avoid destructive wireless interference among the energy harvesting and data transmission phases of the system. Moreover, to fully reap the RIS passive beamforming gain for energy harvesting, the transmit beamforming at the source needs to be jointly designed with the RIS phase-shifts.

\begin{figure}[t]
\centering   
\includegraphics[width=0.5\textwidth]{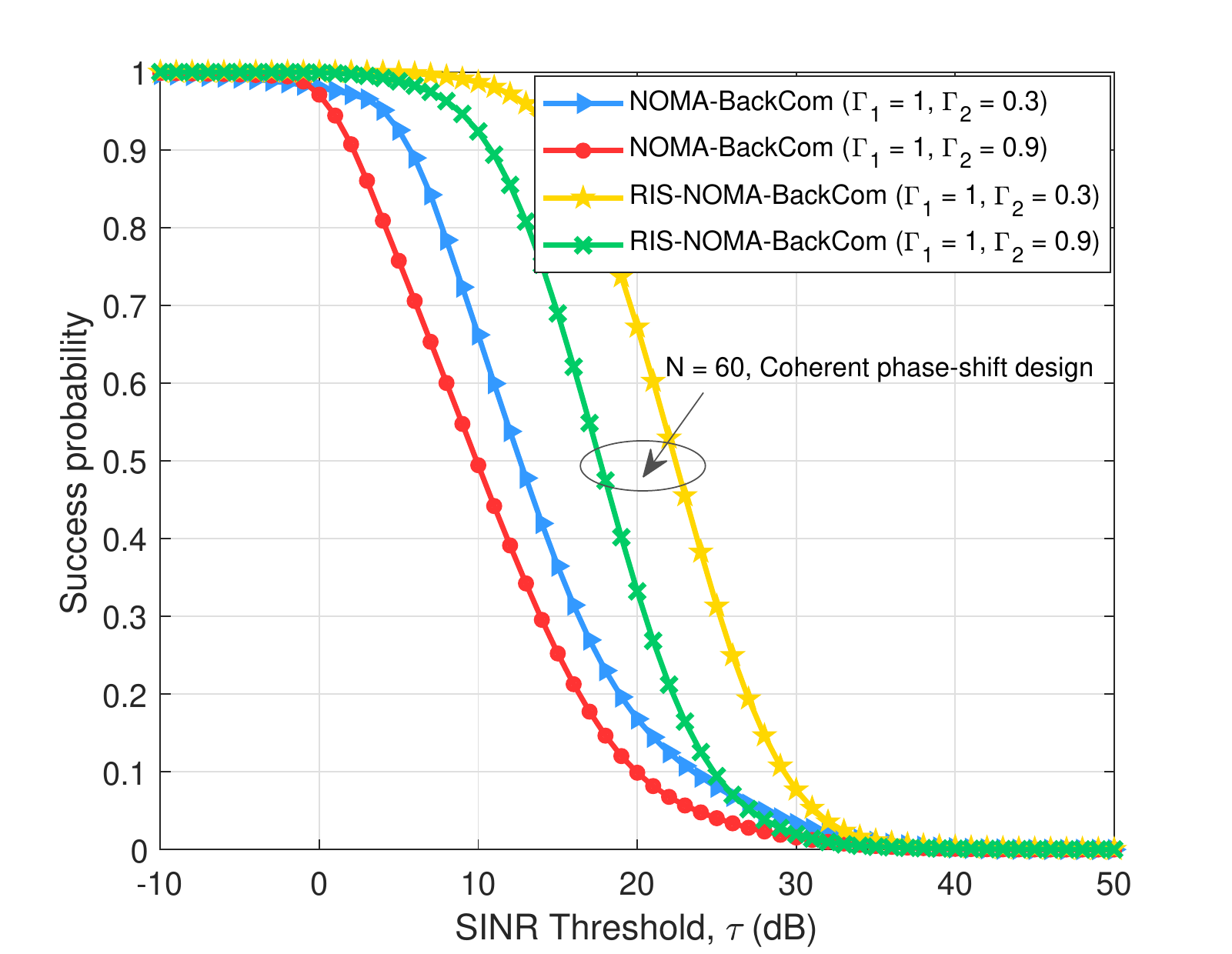}
\caption{Success probability versus the SINR threshold for different values of reflection coefficients, and CE transmit power of $35$ dBm.}\label{fig.4}
\end{figure}

\section{RIS-Assisted NOMA-Enhanced BackCom System: A Case Study}
To support massive connectivity of an increasing number of IoT devices, non-orthogonal multiple access (NOMA) is a promising multiple access candidate due to its capability of exploiting available resources efficiently. In NOMA-enhanced BackCom systems, multiple devices are multiplexed over the same orthogonal resource (e.g., time, frequency, code) block but with different backscattered power levels. With this, NOMA promotes high spectral efficiency, massive connectivity, lower latency, and better user fairness over orthogonal multiple access (OMA) techniques~\cite{Shahzeb_NOMA_BackCom_Mag}. For the emerging 6G IoT networks, the performance of NOMA-enhanced BackCom systems can be improved by incorporating RIS. Therefore, in this section, we the integration of RIS into a NOMA-enhanced bistatic BackCom system.


\vspace{-0.05cm}
\subsection{System Model}
As illustrated in Fig.~\ref{fig.3}, we consider an RIS-assisted NOMA enhanced bistatic BackCom system, which consists of a CE, \(K\) BDs, a reader, and an RIS, where \(N\) reflecting elements only adjusts the phases of the incident signals. The RIS is deployed to enhance the channel gains over both the excitation and backscattering phases of the bistatic BackCom system, where the CE transmits the CW signal to the BDs, while each BD modulates its information over the incident CW signal and backscatters it to the reader. We assume that each of the CE, BR, and BDs are equipped with a single antenna. 
\begin{figure}[t]
\centering   
\includegraphics[width=0.5\textwidth]{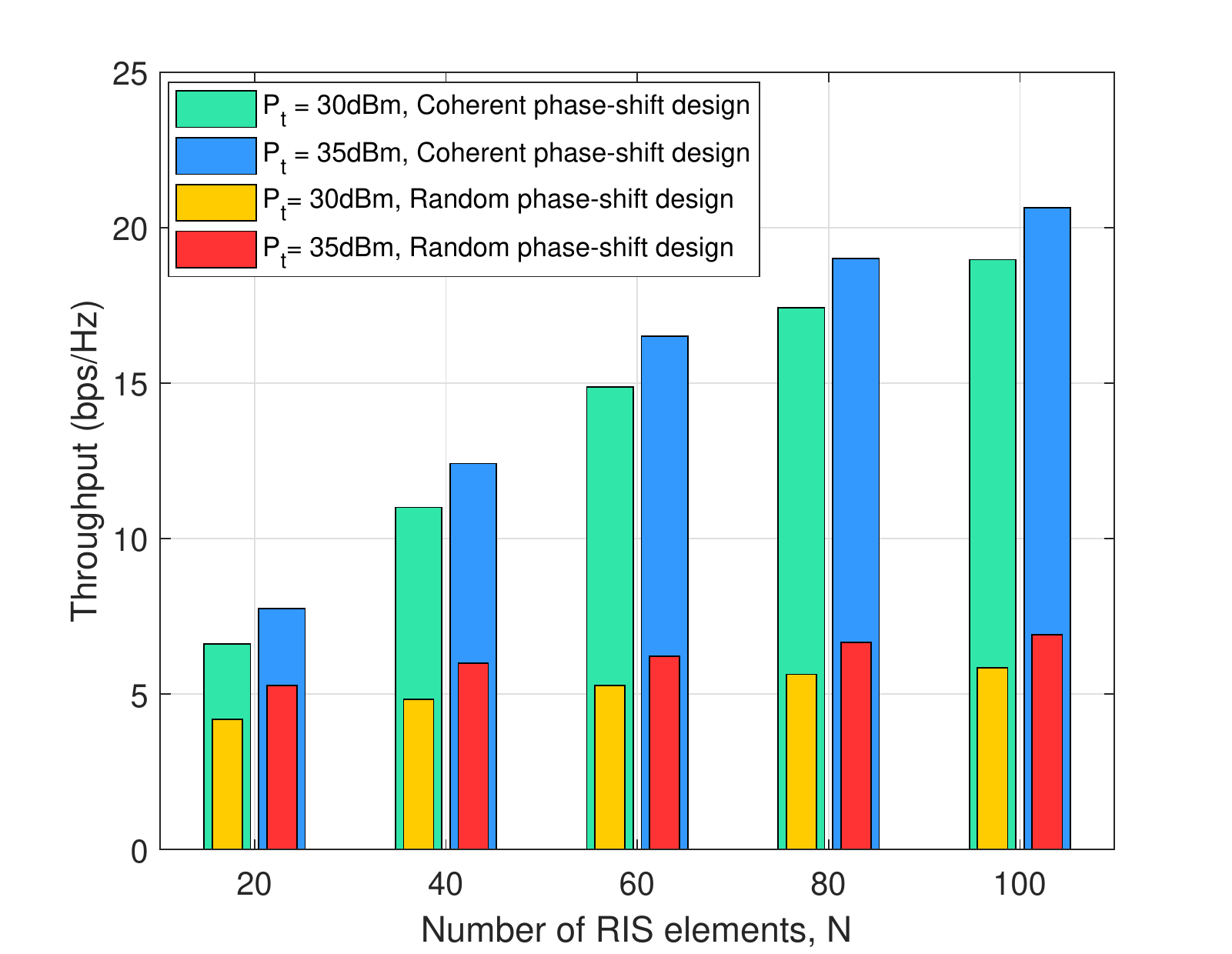}
\caption{Throughput for coherent and random phase-shift designs of RIS with varying number of RIS elements, CE transmit, and SINR threshold of $15$ dB.}\label{fig.5}
\end{figure}

\subsubsection{\textbf{BackCom Model}}

We assume that the CE transmits the CW signal almost all the time, while each BD has two states: (i) sleep state, where the BDs harvest energy from the incident CW signals and store the harvested energy in the battery to power the circuitry and perform sensing operations; (ii) active state, where the BDs backscatter the modulated signal to the BR by intelligently changing their load impedance. To modulate the signal, we adopt the binary phase shift keying (BPSK) modulation. Therefore, each impedance set can generate two reflection coefficients with the same magnitude, denoted as \(\Gamma_i\), but different phase-shifts, i.e., \(0\degree\) and \(180\degree\). We consider $2$-BD NOMA multiplexing, therefore, each BD is equipped with two impedance sets, with the reflection coefficients \(\Gamma_1\) and \(\Gamma_2\), where \(\Gamma_1\) \(>\) \(\Gamma_2\)~\cite{Ahsen_waleed_NOMA_BackCom}.

\subsubsection{\textbf{Frame-Based Transmission Protocol}}
As illustrated in Fig.~\ref{fig.3}, we propose a frame-based transmission protocol that integrates $2$-BD NOMA multiplexing and time-division-multiple-access (TDMA) with \(K\) training slots, for the \emph{training phase}, and \(M\) transmission slots, for the \emph{transmission phase}. In each training slot, only one BD backscatters the CW signal with the same reflection
coefficient, with all other BDs in the sleep state, and the BR
estimates the BD’s cascaded channel, i.e., which is the sum of the direct and the RIS reflected channel. Then, based on the BR’s received power levels, the BDs are sorted decreasingly and categorized sequentially into
two power groups, namely the higher-power BDs and the lower-power BDs.  Each higher-power BD is randomly grouped with a lower-power BD to form a cluster. In each transmission slot, only one cluster performs the NOMA transmission, with higher- and lower-power BDs switch to reflection coefficient \(\Gamma_1\) and \(\Gamma_2\), respectively, while the rest of the clusters remain in the sleep state.
\vspace{-0.6cm}
 
\subsection{Performance Evaluation}
In this subsection, we demonstrate the performance gains achieved by integrating RIS in NOMA-enhanced bistatic BackCom system. We assume that the RIS reflected channels, i.e., channels between the RIS and the CE,
BDs and BR, undergo Rician fading, with Rician factor $3$ dB and path loss exponent $2.4$; while all other channels experience Rayleigh fading with path loss exponent $3$. All wireless channels are assumed to be mutually independent and perfectly estimated by the BR. As in~\cite{MBC_BBC_Tx_min}, the CE, BR, and RIS are located
at \([0, 0]\), \([100,0]\), and \([20, 0]\) m, respectively, and the BDs are located between \([5,0]\) and \([95,0]\) m. The noise power at the BR is considered to be $-90$ dBm. The simulation results for a single cluster case are presented as follows.

\subsubsection{\textbf{Performance Improvement and the Effect of Reflection Coefficients}}
In Fig.~\ref{fig.4}, we evaluate the performance of the proposed RIS-assisted system, i.e., RIS-NOMA-BackCom, against the success probability, defined as the probability by which the received signal-to-interference plus noise ratio (SINR) is greater than the specified threshold, \(\tau\), required for successful decoding of each BD's signal. It can be observed that for any combination of the reflection coefficients, the RIS-assisted system outperforms the no-RIS counterpart, i.e., NOMA-BackCom. Moreover, better performance can be achieved by reducing the value of \(\Gamma_2\). This is because the smaller value of \(\Gamma_2\) reduces the interference from the weaker signal, thus increasing the chances of the stronger signal to be decoded successfully. However, the best performance can be achieved by setting the optimal values for \(\Gamma_1\) and \(\Gamma_2\).

\subsubsection{\textbf{Effect of the RIS Elements and Phase-shift Designs}}
The performance of RIS-assisted system highly depends on the RIS phase-shift design and the number of RIS elements. In this regard, Fig.~\ref{fig.5} compares the cluster throughput for the coherent and random phase-shift designs. First, it can be observed that for a given number of RIS elements and CE transmit power, the coherent phase-shift design, for which the RIS phase-shifts are matched with the phases of the fading channels~\cite{Zhiguo_Ding_IRS-NOMA}, outperforms the random phase-shift design. Second, the throughput scales up with the increase in the RIS elements. For instance, for $30$ dBm CE power and coherent phase-shift design, throughput is $11$ bps/Hz for $40$ RIS elements, while this value increases to about $15$ bps/Hz for $60$ RIS elements. From here, we can conclude that RIS passive reflection adds power gain, which can be either utilized to improve the BackCom system throughput or reduce the CE power consumption.

\subsubsection{\textbf{Energy Harvesting Performance}}
\begin{figure}[t]
\centering   
\includegraphics[width=0.5\textwidth]{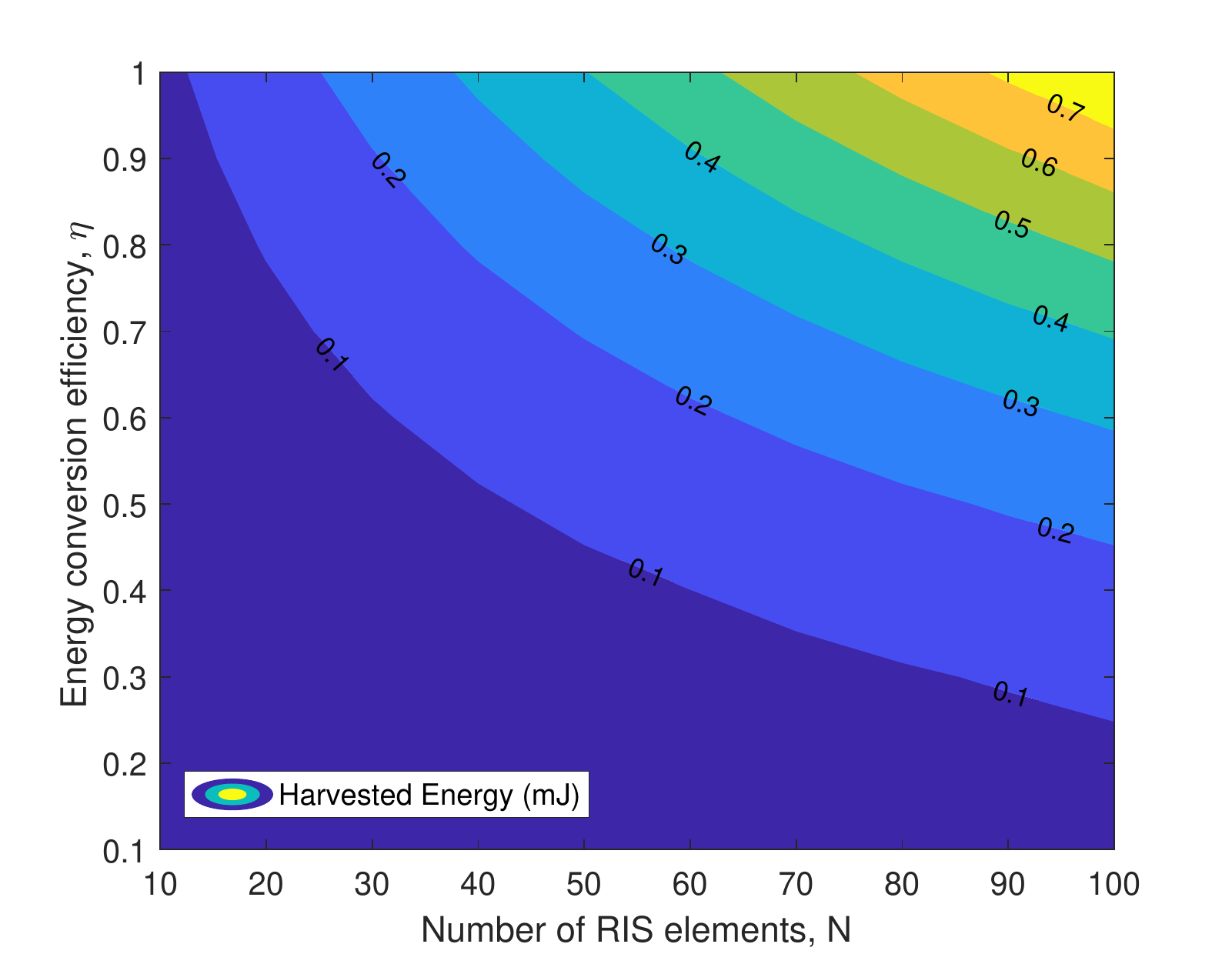}
\caption{Harvested energy for varying number of RIS elements and energy conversion efficiency with CE transmit power of $30$ dBm.}\label{fig.6}
\end{figure}

To evaluate the energy harvesting performance, Fig.~\ref{fig.6} highlights the impact of the increasing number of RIS elements on the harvested energy in the sleep state. It can be observed that for given energy conversion efficiency, the harvested energy increases with the increase in RIS elements. For instance, for $20$\% energy conversion efficiency, the harvested energy increases by $0.15$ mJ when RIS elements increase from $50$ to $100$. Therefore, in practical BackCom systems with low energy conversion efficiency, greater energy can be harvested by deploying RIS with massive number of elements. 

The aforementioned results clearly depict the significance of incorporating RIS into NOMA-enhanced BackCom systems, thus unveiling it as a potential candidate to support the demands for 6G IoT networks.

\section{Challenges and Research Opportunities}
Despite so many promising aspects, as a new-born thing, RIS-BackCom also raises up some challenges and research opportunities which are discussed as follows.

\subsection{Channel Estimation for RIS-BackCom Systems}
The enormous passive beamforming gains brought by RIS highly depend on the availability of CSI, which is quite challenging to acquire since passive RIS elements and backscatter devices lack signal processing capabilities. Thus, only the composite channel, which is the product of the forward and backward channels, can be estimated at the backscatter receiver. This makes the channel estimation schemes available for conventional RIS-assisted systems inapplicable. Therefore, novel schemes need to be developed for channel estimation of RIS-BackCom systems, while accounting for the practical constraints.

\subsection{Optimal Configuration and Resource Allocation}
The RIS configuration highly impact the overall performance of RIS-BackCom systems. The optimal configuration of RIS phase-shifts for both the excitation and backscattering phases renders high computational complexity and latency. Therefore, efficient algorithms are required to control the phase-shifts in a timely manner, according to the dynamics of the propagation environment. In addition, low-complexity resource allocation algorithms need to be developed for RIS-BackCom systems.

\subsection{Machine Learning-empowered RIS-BackCom }
Owing to the limited CSI and non-convex nature of optimization variables, machine learning (ML) techniques are quite appealing for RIS-BackCom systems, due to their learning capability and large search space. Compared to conventional optimization and signal processing techniques, ML can achieve a high degree in all respects of RIS-BackCom, including environmental sensing, channel estimation, phase-shift design, and resource allocation. This, thus calls for efficient ML-based algorithms to fully reap the potentials of RIS-BackCom systems.

\subsection{mmWave-based RIS-BackCom}
With the capability to exploit the under-utilized immense bandwidth at high-frequency spectrum, mmWave has been deemed as a serious candidate to host the high data rate communications. Nevertheless, the vulnerability of mmWaves to signal blockages renders it insufficient on its own to meet the growing demands of wireless communication. Fortunately, RIS has the capability to mitigate the limitations of mmWave systems by introducing effective additional paths. In this context, mmWave-based RIS-BackCom is worth investing in for realizing future IoT networks.

\subsection{Integrating Multiple RISs into BackCom Systems}
In future large-scale networks, multiple RISs are anticipated to be deployed in the BackCom systems to provide ultra-high performance, since multiple RISs open up a door for harvesting larger beamforming gains. However, the deployment of multiple RISs incurs high design and implementation complexity. Therefore, the cumulative impact of the operation of multiple RISs is worth investigating, by considering their spatial distribution.

\section{Conclusion}
In this article, we explored the integration of RISs into BackCom systems for enabling 6G IoT networks. Specifically, we first outlined the fundamentals of RIS technology to develop an understanding of RIS in BackCom. We introduced four different variants of RIS-BackCom and identified the potential improvement that can be achieved by incorporating RISs into BackCom systems. In addition, a case study for an RIS-assisted NOMA-enhanced BackCom system is presented. Our numerical results clearly demonstrated the performance gains brought by RIS in terms of success probability, throughput, transmit power consumption, and harvested energy. Finally, to provide useful guidance for future research, we indicated the crucial challenges and promising research directions for realizing RIS-BackCom systems. 

\input{output.bbl}

\vspace{-2cm}
\begin{IEEEbiographynophoto}{Sarah Basharat}(sbasharat.msee19seecs@seecs.edu.pk) received her B.E. and M.S. degrees in electrical engineering from National University of Sciences and Technology (NUST), Pakistan, in 2019 and 2021, respectively. Her research interests include B5G and 6G communication, non-orthogonal multiple access (NOMA), backscatter communication (BackCom), and reconfigurable intelligent surfaces (RISs).

\end{IEEEbiographynophoto}
\vspace{-2cm}
\begin{IEEEbiographynophoto}{Syed Ali Hassan} [S’08, M’11, SM’17] (ali.hassan@seecs.edu.pk) received his Ph.D. in electrical engineering from Georgia Tech, Atlanta, in 2011, his M.S. in mathematics from Georgia Tech in 2011, and his M.S. in electrical engineering from the University of Stuttgart, Germany, in 2007, and a B.E. in electrical engineering (highest honors) from the National University of Sciences and Technology (NUST), Pakistan, in 2004. Currently, he is working as an associate professor at NUST, where he is heading the IPT research group, which focuses on various aspects of theoretical communications. 
\end{IEEEbiographynophoto}

\vspace{-2cm}
\begin{IEEEbiographynophoto}{Aamir Mahmood}[M’18, SM’19] (aamir.Mahmood@miun.se) is an assistant professor of communication engineering at Mid Sweden University, Sweden. He received the M.Sc. and D.Sc. degrees in communications engineering from Aalto University School of Electrical Engineering, Finland, in 2008 and 2014, respectively. He was a research intern at Nokia Researcher Center, Finland and a visiting researcher at Aalto University during 2014-2016. His research interests include network time synchronization, resource allocation for URLLC, and RF interference/coexistence management.
\end{IEEEbiographynophoto}
\vspace{-2cm}
\begin{IEEEbiographynophoto}{Zhiguo Ding}[S’03, M’05, SM’17, F’20] (zhiguo.ding@manchester.ac.uk) is currently a Professor at the University of Manchester. Dr Ding research interests are 5G networks, signal processing and statistical signal processing. He has been serving as an Editor for IEEE TCOM, IEEE TVT, and served as an editor for IEEE WCL and IEEE CL. He received the EU Marie Curie Fellowship 2012-2014, IEEE TVT Top Editor 2017, 2018 IEEE COMSOC Heinrich Hertz Award, 2018 IEEE VTS Jack Neubauer Memorial Award, and 2018 IEEE SPS Best Signal Processing Letter Award.
\end{IEEEbiographynophoto}
\vspace{-2cm}
\begin{IEEEbiographynophoto}{Mikael Gidlund}[M’98, SM’16] (mikael.gidlund@miun.se) is a professor of computer engineering at Mid Sweden University, Sweden. He has worked as Senior Principal Scientist and Global Research Area Coordinator of Wireless Technologies, ABB Corporate Research, Sweden, Project Manager and Senior Specialist with Nera Networks AS, Norway, and Research Engineer and Project Manager with Acreo AB, Sweden. His current research interests include wireless communication and networks, wireless sensor networks, access protocols, and security. 
\end{IEEEbiographynophoto}

\end{document}